\documentclass{aa}

\usepackage{epsfig}

\begin{document}

\title{A precessing accretion disc in the intermediate polar XY Ari?}
 
\titlerunning{Precessing accretion disc in XY Ari?}

\author{A.J. Norton\inst{1} \and K. Mukai\inst{2} }

\authorrunning{Norton \& Mukai}

\offprints{A.J. Norton}

\institute{Department of Physics and Astronomy, The Open University, 
	Walton Hall, Milton Keynes MK7 6AA, U.K. \\
	\email{A.J.Norton@open.ac.uk,}
 \and
	CRESST and X-ray Astrophysics Laboratory NASA/GSFC, Greenbelt, 
        MD 20771, USA {\it and} Department of Physics, University of Maryland,
        Baltimore county, 1000 Hilltop Circle, Baltimore, MD 21250, USA \\
	\email{mukai@milkyway.gsfc.nasa.gov}
}

\date{Received ???;
      accepted ???}

\abstract
{XY Ari is the only intermediate polar to show deep X-ray eclipses of its 
white dwarf. Previously published observations with \begin{em}Ginga\end{em}
and \begin{em}Chandra\end{em} have also revealed a broad X-ray orbital 
modulation, roughly antiphased with the eclipse, and presumed to be due to 
absorption in an extended structure near the edge of an accretion disc. The
X-ray pulse profile is generally seen to be double-peaked, although a 
single-peaked pulse was seen by \begin{em}RXTE\end{em} during an outburst in 
1996.}
{We intended to investigate the cause of the broad orbital modulation in 
XY Ari to better understand the accretion flow in this system and other 
intermediate polars.}
{We observed XY Ari with \begin{em}RXTE\end{em} and analysed previously
unpublished archival observations of the system made with 
\begin{em}ASCA\end{em} and \begin{em}XMM-Newton\end{em}. 
These observations comprise six separate visits and span about ten years.}
{The various X-ray observations show that the broad orbital modulation 
varies in phase and significance, then ultimately disappears entirely in the 
last few years. In addition, the X-ray pulse profile shows variations in 
depth and shape, and in the recent \begin{em}RXTE\end{em} observations 
displays no evidence for changes in hardness ratio.}
{The observed changes indicates that both the pulse profile and 
the orbital modulation are solely due to geometrical effects at the time
of the \begin{em}RXTE\end{em} observations, rather than phase-dependent 
variations in photoelectric absorption as seen previously. We suggest that 
this is evidence for a precessing, tilted accretion disc in this system.
The precession of the disc moves structures out of our line of sight both
at its outer edge (changing the orbital modulation) and at its inner edge
where the accretion curtains are anchored (changing the pulse profile).}


\maketitle

\section{Introduction}

XY Ari is an intermediate polar (IP), namely a semi-detached interacting 
asynchronous binary star in which a magnetic white dwarf with a 
surface field strength of order a few megagauss, accretes 
material from a Roche lobe-filling, usually main sequence dwarf, companion. 
The accretion flow from the donor star generally forms a truncated 
accretion disc whose inner edge is set by the white dwarf's magnetospheric
 radius. Here the material attaches to the magnetic field lines before 
following them towards the white dwarf magnetic poles. Arc-shaped accretion 
curtains are believed to form, standing above the white dwarf surface. 
Towards the base of these curtains, the accretion flow undergoes a strong 
shock, below which material settles onto the white dwarf, releasing X-ray to 
optical emission. Since the magnetic axis is offset from the spin axis of the 
white dwarf, this gives rise to the defining characteristic of the class, 
namely X-ray (and usually optical) emission pulsed at the white dwarf spin 
period. Around 30 confirmed IPs are now recognised, with a further 20 or so 
candidate systems having been proposed\footnote{See 
http://asd.gsfc.nasa.gov/Koji.Mukai/iphome/iphome.html for an on-line
catalogue of systems.}. Comprehensive reviews of various 
aspects of their behaviour are given by  Patterson (1994), Warner (1995) 
and Norton, Wynn \& Somerscales (2004).

The pulse profiles of IPs are variously seen to be 
single-peaked (e.g. AO~Psc, V1223~Sgr, TX Col, EX Hya, etc), double-peaked
(YY Dra, V405 Aur, GK Per, etc), or displaying more complex shapes 
(FO Aqr, BG CMi, PQ Gem, etc). It is apparent that most of these profile
shapes are due to a combination of emission from {\em both} upper and lower
accreting magnetic poles of the white dwarf. Depending on the geometry
and the relative column density parallel to and perpendicular to the 
white dwarf surface, either a single-peaked or double-peaked profile can
result (Norton et al 1999). Introducing asymmetries
between the two poles can yield more complex profiles (Beardmore et al 1998).
During outbursts, GK Per and XY Ari have each been seen to develop a
single-peaked pulse profile (Watson, King \& Osborne 1985; Hellier, 
Beardmore \& Mukai 1997). This is interpreted as arising when the lower
pole becomes hidden by the inner extension of the outbursting accretion
disc as it forces the magnetosphere inwards (Hellier et al 1997; Hellier,
Harmer \& Beardmore 2004).

Around half of the known IPs display orbitally modulated X-ray flux (Parker,
Norton \& Mukai 2005). The cause of this modulation is generally believed 
to be local absorption of X-rays in an extended structure at the 
outer edge of the accretion disc, caused by the impact of the stream from
the inner Lagrangian point with the disc itself. Absorption at the outer edge 
of a disc is the accepted explanation for X-ray orbital modulations seen in 
Low Mass X-ray Binaries too. In our systematic study of orbital modulation in
sixteen IPs, covering thirty observations with {\em RXTE} and {\em ASCA}, we 
noted that the presence of
orbital modulation can appear and disappear on timescales of years or months in
an individual system. This was particularly apparent in AO Psc and V1223 Sgr, 
possibly because these have been some of the best studied systems over the
years so there is a wealth of observational data sets for each system,
capturing its behaviour at many different epochs. 

\section{The history of XY Ari}

XY Ari was originally discovered as the X-ray source 1H~0253+193 by the 
{\em Einstein} satellite (Halpern \& Patterson 1987). Subsequent short
{\em Ginga} observations in July 1987 and January 1989 discovered a coherent 
X-ray pulsation with a period of 206~s and a double-peaked profile (Takano 
et al 1989; Koyama et al 1991), suggesting its identification as an IP 
(Patterson \& Halpern 1990). Further, longer observations 
with the same satellite in August 1989 discovered eclipses with a period of 
6.06~hr (Kamata, Tawara \& Koyama 1991), so establishing the orbital period 
of the system and making it the first (and so far only) IP to exhibit deep 
X-ray eclipses. 

Since XY Ari lies behind the molecular cloud MBM12, no optical counterpart 
can be seen, but the system is detected in the infrared, where an ellipsoidal 
modulation is apparent (Allan, Hellier \& Ramseyer 1996). Modelling of this 
lightcurve allows determination of the system parameters. In particular the 
mass ratio is determined to lie in the range $0.43 < q < 0.71$ and the 
system inclination angle in the range $80^{\circ} < i < 87^{\circ}$. Infrared 
spectroscopy of XY Ari has allowed its distance to be estimated as 
$270 \pm 100$~pc (Littlefair, Dhillon \& Marsh 2001).

The eclipsing nature of XY Ari is important, as accurate timing of the 
eclipse ingress and egress allows a determination of the size of the X-ray 
emitting region on the WD surface. Hellier (1997) used a series of short 
{\em RXTE} observations, spanning many eclipse egresses, to show that the 
emitting region has a size smaller than 0.002 of the WD surface area, and 
that the accretion region itself may wander over a larger area, of fractional 
size less than 0.01 of the WD surface, on a timescale comparable to the 
orbital period. Hellier (1997) also used this study of the eclipse to constrain
the upper limit on the inclination angle further, namely $i < 84^{\circ}$.

A further interesting aspect of the behaviour of XY Ari was revealed by 
these same short {\em RXTE} observations, namely that the source exhibitted
an outburst in July 1996, during which the mass accretion rate increased 
substantially (Hellier, Mukai \& Beardmore 1997). The outburst was consistent 
with being due to a thermal-viscous disc instability, as in conventional 
dwarf novae, although alternatives, such as an enhanced mass transfer rate 
from the secondary, could not be ruled out.

Also of interest for the present study is the fact that the system displays
a broad, sinusoidal X-ray orbital modulation, on which the deep X-ray eclipse 
is superimposed. In the {\em Ginga} observations from 1989, Kamata et al (1991)
show that the minimum of the broad modulation lags the 
narrow eclipse by about 0.45 in orbital phase, with the eclipse occurring
just {\em after} the maximum of the broad modulation. They further demonstrate
that the broad modulation is antiphased with the X-ray hardness ratio, 
implying that the cause of the modulation is photoelectric absorption.
The only other published X-ray observations of XY Ari of sufficient duration 
to observe the full orbital cycle are those by {\em Chandra} in July/August
2000, reported by Salinas \& Schlegel (2004). Although only covering
parts of four orbital cycles, and of lower signal-to-noise than the {\em Ginga}
observations, these lightcurves too show the broad orbital modulation.
The phasing here is such that the broad orbital minimum lags the deep
eclipse by $\sim 0.65 - 0.75$ in orbital phase, with the eclipse occurring
just {\em before} the maximum of the broad X-ray modulation. Based on 
just these two observations therefore, there is already evidence for
variability in the broad X-ray modulation component in this system.

\section{Observations and data reduction} 

\subsection{New {\em RXTE} observations}

The new data presented here were obtained with the PCA instrument on the 
{\em RXTE} satellite. The observation sequence comprises eight segments, 
each of about 6 hours duration on the 19th, 20th, 23rd and 24th November 2005 
and on the 7th, 8th, 10th and 11th January 2006. Each of the eight segments 
comprises data from four satellite orbits, resulting in typically 
$\sim 11$~ksec of usable on-source data from each. Various combinations of 
the 5 proportional counter units of the PCA were active during each 
observation segment. Neglecting PCU0, for which the background subtraction
is problematical since the loss of the propane layer, data from 
either 1, 2 or 3 PCUs were available at any time.

Data were extracted using standard criteria from the top layer only of each
PCU, to make lightcurves with 16~sec time resolution, in each of four 
energy ranges, namely 2--4 keV, 4--6 keV, 6--10 keV and 10--20 keV. 
Background lightcurves were generated using the latest {\em RXTE} background
models and subtracted from the on source lightcurves. To account for the 
different number of PCUs active at any time, the resulting count rates
were simply scaled by the number of active PCUs, to yield counts per 
second per PCU. 

The power spectra of the resulting lightcurves, both from the separate 
November and January visits and from the combined observation, 
are dominated by signals at high harmonics of the orbital frequency, due to 
the narrow X-ray  eclipse. There is no signal at the fundamental orbital 
frequency in the power spectra of any of the lightcurves. Signals at the 
spin frequency of the white dwarf, and its first harmonic, are only 
marginally significant.

\begin{figure*}[ht]
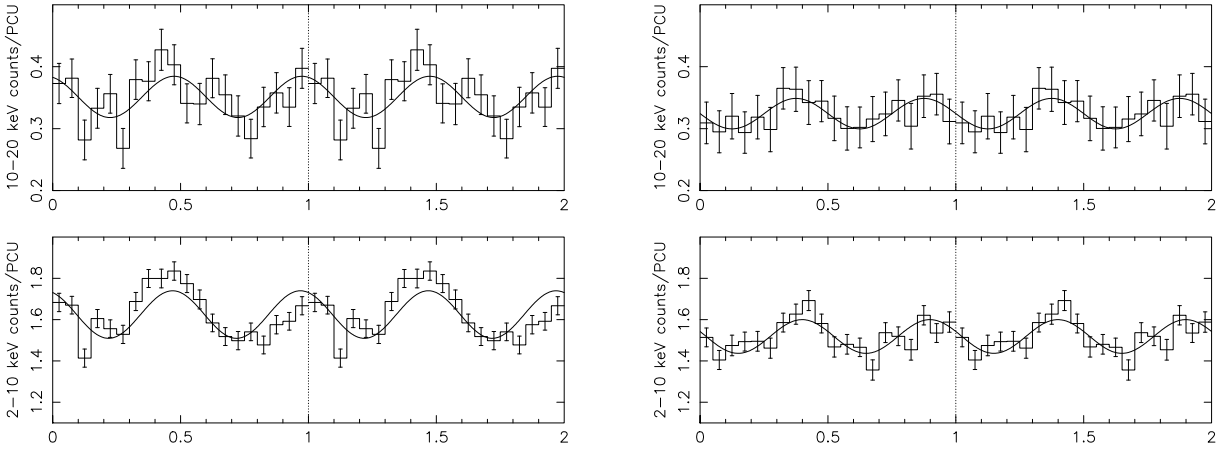

\epsfig{file=7761fig1a.eps, angle=0, width=8.5cm}
\epsfig{file=7761fig1b.eps, angle=0, width=8.5cm}
\caption{The {\em RXTE} X-ray lightcurves of XY Ari from November 2005 
(left) and January 2006 (right) folded at the 206~s
white dwarf spin period. In the upper panel the energy range is 
10--20 keV, whilst in the lower panel it is 2--10 keV. The sinusoids 
overplotted are the best fit to any double peaked pulse profile.}
\end{figure*}

\begin{figure*}[ht]
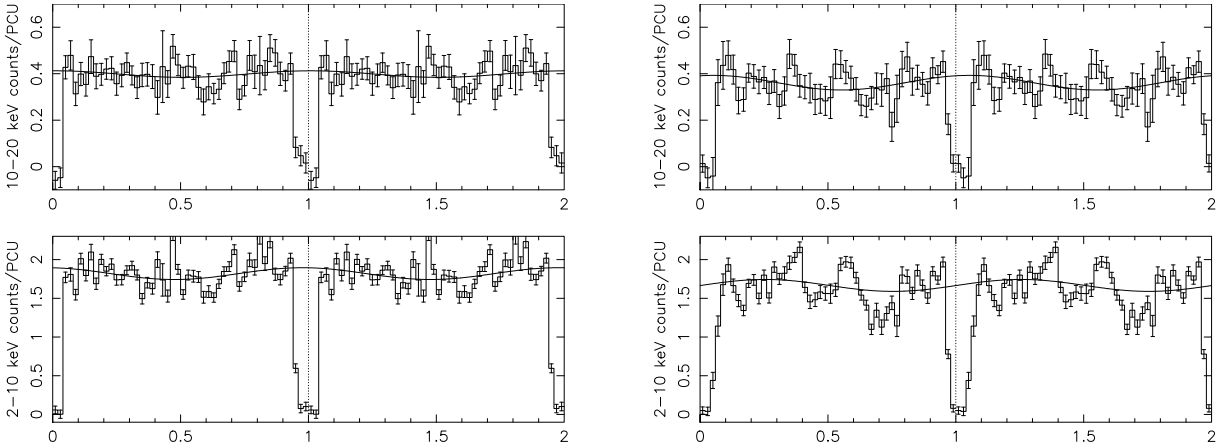

\epsfig{file=7761fig2a.eps, angle=0, width=8.5cm}
\epsfig{file=7761fig2b.eps, angle=0, width=8.5cm}
\caption{The {\em RXTE} X-ray lightcurves of XY Ari from November 2005 (left)
and January 2006 (right) folded at the 6.06~hr
orbital period of the binary. In the upper panel the energy range is 
10--20 keV, whilst in the lower panel it is 2--10 keV. The sinusoids 
overplotted are the best fit to any broad modulation, excluding the eclipse 
itself.}
\end{figure*}

The lightcurves for the two halves of the observation (from November 2005 
and January 2006), folded at the 206~s 
white dwarf spin period are shown in Figure 1. Although four energy 
ranges were extracted, as noted above, we show here the combined 2--10~keV
lightcurve for ease of comparison with other datasets presented in later
sections. The familiar double 
peaked pulse profile is seen in each energy band, albeit with a rather
poor signal-to-noise ratio. Phase zero is arbitrary in these profiles. We 
fitted a sinusoid to these data with a period
equal to half the white dwarf spin period, but with the amplitude, mean level 
and phase of minimum flux as free parameters in each case. This yielded the 
curves which are over-plotted on Figure 1. The depths of this modulation are 
listed in Table 1 and defined as the peak-to-peak range of the sinusoid 
divided by the maximum value. The modulation depths here are all rather low, at
around $10\%-15\%$ at all energies at both epochs. The constancy 
of the modulation depth is further demonstrated by the fact that the 
(2--6 keV / 6--10 keV) hardness ratio is constant when folded at the spin 
period. 

The {\em RXTE} lightcurves folded at the 6.06~hr orbital period are shown in
Figure 2. As for the pulse profiles, although lightcurves were extracted
in four separate energy bands, we show here the combined 2--10~keV orbital
profile for ease of comparison with other datasets presented in later 
sections. The narrow eclipse is clearly present, with the X-ray flux 
virtually extinguished for $\sim 0.15$ of the orbital cycle, but 
the broad orbital modulation seen in previous observations has 
essentially disappeared. As for the pulse profiles, we fitted a sinusoid to 
these data (excluding the eclipse itself), with the amplitude, mean level and 
phase of minimum flux as free parameters in each case. This yielded the 
curves which are over-plotted on Figure 2; the depths of the broad modulation 
are listed in Table 2. For these {\em RXTE} observations, the modulation 
depth in all energy bands is consistent with zero at the $\sim 2\sigma$ level.

\begin{table*}[h]
\caption{The double-peaked modulation depth of the pulse profile in various 
X-ray observations of XY Ari}
\begin{tabular}{lllrlrlrlrlrr} \hline
Mission & Date & \multicolumn{2}{c}{1--2 keV} & \multicolumn{2}{c}{2--4 keV} & \multicolumn{2}{c}{4--6 keV} & \multicolumn{2}{c}{6--10 keV} & \multicolumn{2}{c}{10--20 keV} & Ref \\ \hline
{\em Ginga} & 1989 Aug & [ $\leftarrow$ & \multicolumn{2}{c}{$\sim 40\%$} & $\rightarrow$ ] & [ $\leftarrow$ & \multicolumn{2}{c}{$\sim 20\%$} & $\rightarrow$ ]& \multicolumn{2}{c}{$\sim 20\%$} & Kamata et al 1991 \\
{\em ASCA}  & 1995 Aug & \multicolumn{2}{c}{single peak}  & [ $\leftarrow$ & \multicolumn{4}{c}{no modulation} & $\rightarrow$] & & & this work \\
{\em ASCA}  & 1996 Jan & \multicolumn{2}{c}{single peak}  & [ $\leftarrow$ & \multicolumn{4}{c}{single peak}  & $\rightarrow$ ] & & & this work \\
{\em RXTE}  & 1996 Jul &      &       & [ $\leftarrow$ & \multicolumn{4}{c}{$\sim 20\%$}  & $\rightarrow$ ] & & & Hellier et al 1997 \\
{\em RXTE}  & 1996 Jul (o/b) &   &    & [ $\leftarrow$ & \multicolumn{4}{c}{single peak}  & $\rightarrow$ ] & & & Hellier et al 1997 \\
{\em Chandra} & 2000 Jul & [ $\leftarrow$ & \multicolumn{6}{c}{single peak} & $\rightarrow$ ] & & & Salinas \& Schlegel 2004 \\
{\em XMM}   & 2000 Aug & \multicolumn{2}{c}{single peak} & \multicolumn{2}{c}{$17\%\pm2\%$} & [ $\leftarrow$ & \multicolumn{2}{c}{$14\%\pm3\%$} & $\rightarrow$ ] & & & this work  \\
{\em XMM}   & 2001 Feb & \multicolumn{2}{c}{single peak} & \multicolumn{2}{c}{$18\%\pm5\%$} & [ $\leftarrow$ & \multicolumn{2}{c}{$14\%\pm3\%$} & $\rightarrow$ ] & & & this work   \\
{\em RXTE}  & 2005 Nov &       &        & \multicolumn{2}{c}{$13\%\pm5\%$} & \multicolumn{2}{c}{$15\%\pm3\%$} &  \multicolumn{2}{c}{$13\%\pm3\%$} & \multicolumn{2}{c}{$17\%\pm6\%$} & this work \\
{\em RXTE}  & 2006 Jan &       &        & \multicolumn{2}{c}{$11\%\pm6\%$} & \multicolumn{2}{c}{$8\%\pm3\%$} & \multicolumn{2}{c}{$11\%\pm3\%$} & \multicolumn{2}{c}{$14\%\pm3\%$} & this work \\
\end{tabular}
\end{table*}

\begin{table*}[h]
\caption{The depth and phase of minimum of the broad orbital modulation 
in various X-ray observations of XY Ari}
\begin{tabular}{lllrlrlrlrlrcr} \hline
Mission & Date & \multicolumn{2}{c}{1--2 keV} & \multicolumn{2}{c}{2--4 keV} & \multicolumn{2}{c}{4--6 keV} & \multicolumn{2}{c}{6--10 keV} & \multicolumn{2}{c}{10--20 keV} & Phase & Ref \\ \hline
{\em Ginga} & 1989 Aug & & & [ $\leftarrow$ & \multicolumn{6}{c}{$\sim 55\%$} & $\rightarrow$ ] & 0.45 & Kamata et al 1991 \\
{\em ASCA}  & 1995 Aug & \multicolumn{2}{c}{$57\%\pm13\%$} & [ $\leftarrow$ & \multicolumn{4}{c}{$16\%\pm5\%$} & $\rightarrow$ ] & & & 0.55 & this work \\
{\em ASCA}  & 1996 Jan & \multicolumn{2}{c}{$36\%\pm10\%$} & [ $\leftarrow$ & \multicolumn{4}{c}{$7\%\pm3\%$} & $\rightarrow$ ]  & & & 0.85 & this work \\
{\em Chandra} & 2000 Jul & [ $\leftarrow$ & \multicolumn{6}{c}{$\sim 40\%$} & $\rightarrow$ ] & & & 0.6 & Salinas \& Schlegel 2004 \\
{\em XMM}   & 2000 Aug & \multicolumn{2}{c}{$63\%\pm9\%$}  & \multicolumn{2}{c}{$25\%\pm7\%$} & [ $\leftarrow$ & \multicolumn{2}{c}{$24\%\pm6\%$} & $\rightarrow$ ] & & & 0.75 & this work \\
{\em XMM}   & 2001 Feb & \multicolumn{2}{c}{$20\%\pm15\%$} & \multicolumn{2}{c}{$13\%\pm8\%$} & [ $\leftarrow$ & \multicolumn{2}{c}{$24\%\pm6\%$} & $\rightarrow$ ] & & & 0.1 &  this work \\
{\em RXTE}  & 2005 Nov & & & \multicolumn{2}{c}{$7\%\pm5\%$} &  \multicolumn{2}{c}{$5\%\pm5\%$} &   \multicolumn{2}{c}{$11\%\pm5\%$} &  \multicolumn{2}{c}{$7\%\pm6\%$} & (undef.) & this work \\
{\em RXTE}  & 2006 Jan & & &  \multicolumn{2}{c}{$9\%\pm6\%$} &  \multicolumn{2}{c}{$7\%\pm6\%$} &   \multicolumn{2}{c}{$12\%\pm6\%$} &  \multicolumn{2}{c}{$16\%\pm7\%$} & (undef.) & this work \\
\end{tabular}
\end{table*}

Fitting a simple model to the average spectra of the November 2005 and 
January 2006 visits reveals that both are adequately fitted 
(with $\chi^2_r =0.77$ in each case)
by a thermal bremsstrahlung model with an iron line. The temperature is 
$kT = 35^{+15}_{-9}$~keV in the first observation and $kT = 
38^{+20}_{-10}$~keV in the second, with hydrogen column densities of 
$N_{\rm H} = 5.3^{+1.1}_{-1.2} \times 10^{22}$~cm$^{-2}$ and  
$N_{\rm H} = 6.7^{+1.4}_{-1.3} \times 10^{22}$~cm$^{-2}$ respectively. 
The spectra are therefore consistent with each other. The 2--10~keV flux
however, shows a slight reduction from 
$1.45\times10^{-11}$~erg~cm$^{2}$~s$^{-1}$ to 
 $1.32\times10^{-11}$~erg~cm$^{2}$~s$^{-1}$, but there are likely to be 
systematic uncertainties of up to $5\%$ on these values due to background
subtraction, judging by the deviations from zero flux at mid eclipse.

\subsection{Archival {\em ASCA} observations}

There are two archival {\em ASCA} observations of XY Ari, neither of which have
previously been published, except in compilation papers examining IP
X-ray spectra (Ezuka \& Ishida 1999; Terada, Ishida \& Makishima 2004). 
They comprise an observation from 1995 August 6
with an exposure of around 35~ksec and another from 1996 January 28 with an
exposure of around 60~ksec. 
We extracted lightcurves in each case from both the SIS and GIS detectors,
in two energy bands. The low energy band corresponds to $\sim 0.7 - 2.0$~keV
whilst the high band is roughly $\sim 2 - 10$~keV.

Background subtracted lightcurves obtained by summing the lightcurves from
the two SIS and two GIS instruments, folded at the white dwarf spin period
in each case are shown in Figure 3. As in Figure 1, phase zero is arbitrary
in these plots. In the August 1995 data there is little
evidence for any modulation above 2~keV, although a possible single-peaked
pulse is visible at low energies. By contrast, in January 1996, there is 
a clear single-peaked pulse profile in both energy bands.

The lightcurves folded at the orbital period are shown in Figure 4. As in 
the case of Figure 2, we show the best fit sinusoid to these folded 
lightcurves (excluding the eclipse) overplotted. Here the depths of the 
modulation, listed in Table 2, are significant at the $\sim 4 \sigma$ level 
in the low energy band, with a more marginal detection in the high energy band.

\begin{figure*}[ht]
\epsfig{file=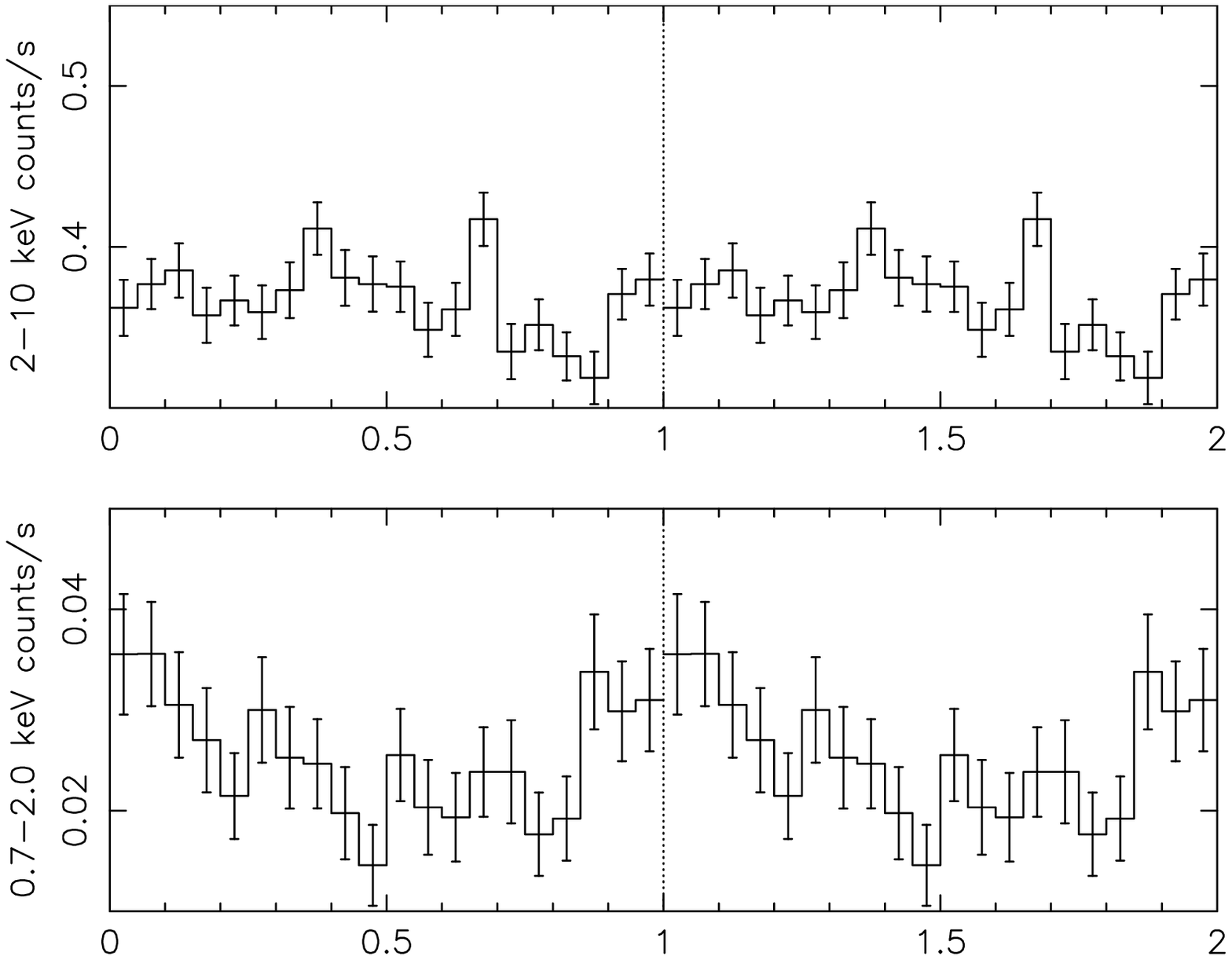, angle=0, width=8.5cm}
\epsfig{file=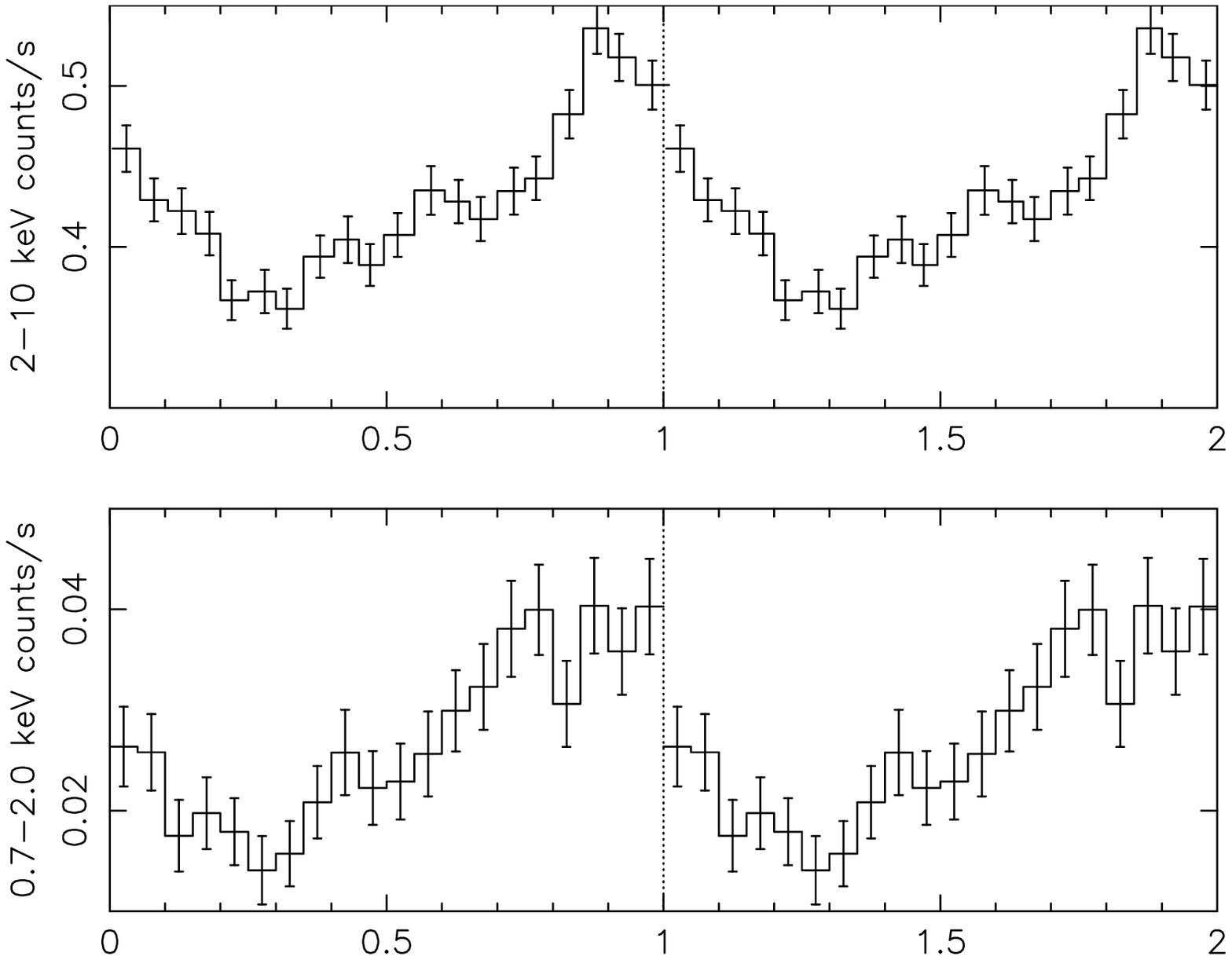, angle=0, width=8.5cm}
\hspace*{-6.5cm}
\caption{The {\em ASCA} X-ray lightcurves of XY Ari from August 1995 (left)
and January 1996 (right) folded at the 206~s white dwarf spin period. The 
energy range of the upper lightcurve is 2--10 keV, whilst that of the lower 
lightcurve is 0.7--2.0 keV.}
\end{figure*}  

\begin{figure*}[ht]
\epsfig{file=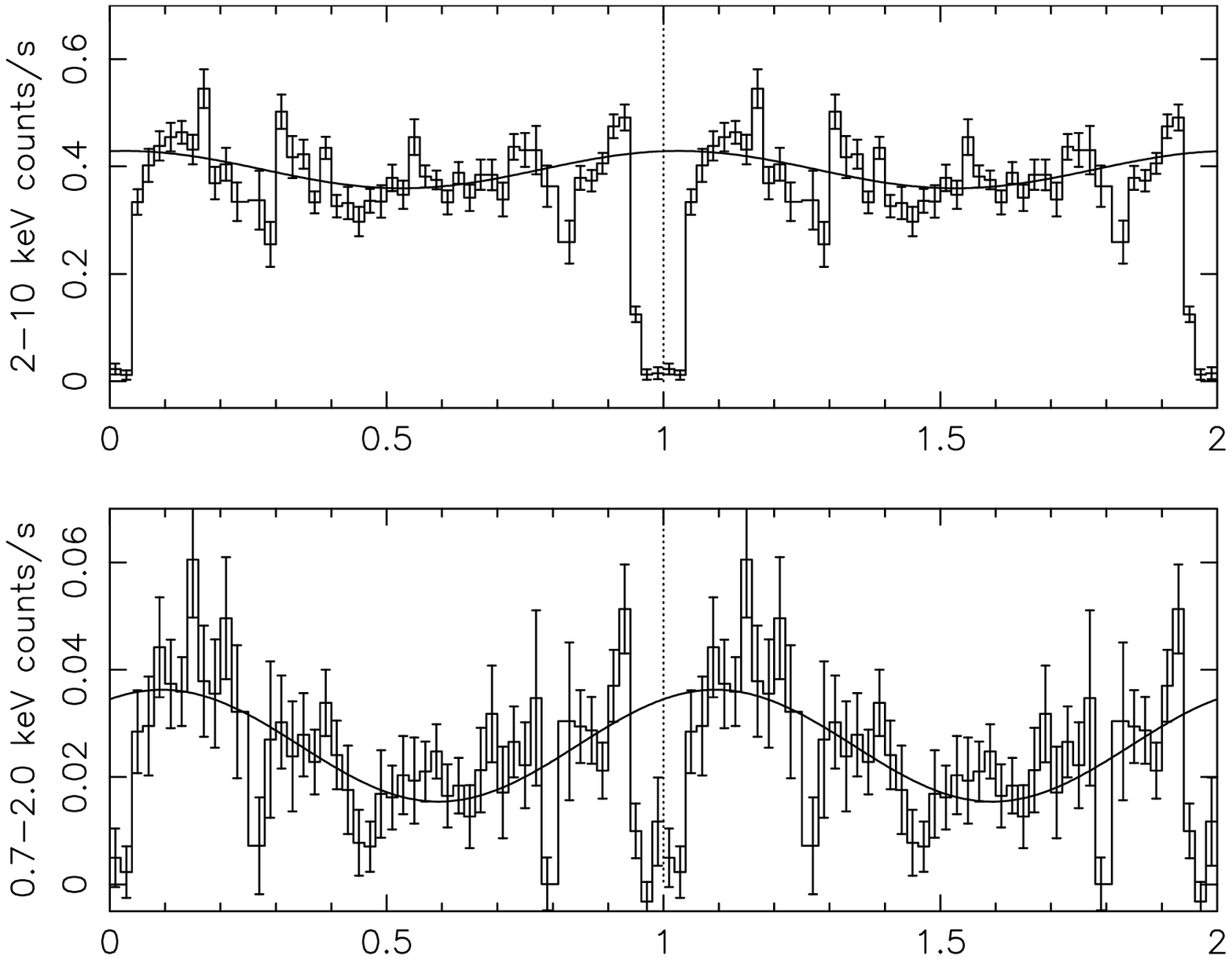, angle=0, width=8.5cm}
\epsfig{file=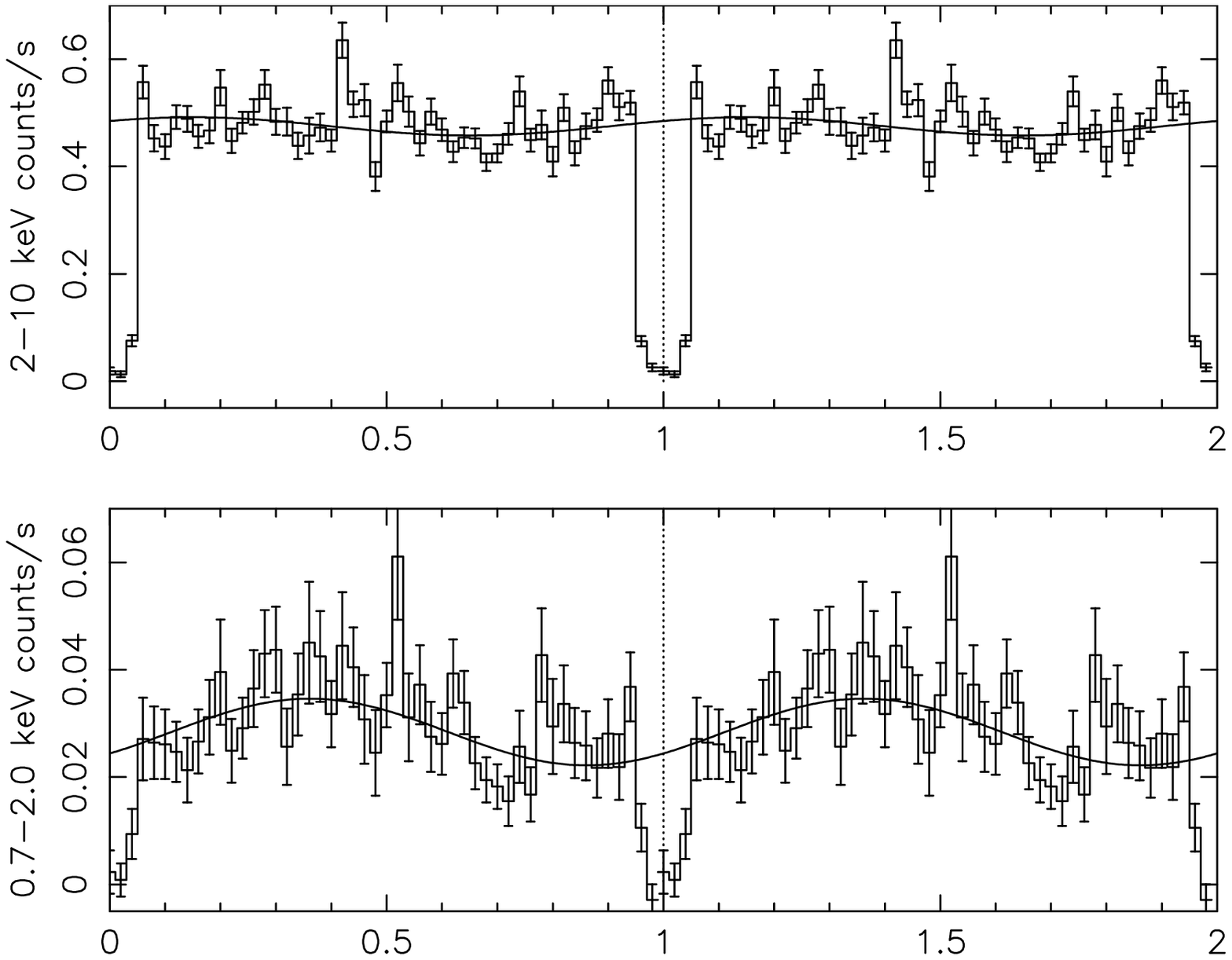, angle=0, width=8.5cm}
\caption{The {\em ASCA} X-ray lightcurves of XY Ari from August 1995 (left)
and January 1996 (right)
folded at the 6.06~hr orbital period of the binary. The energy range
of the upper lightcurve is 2--10 keV, whilst that of the lower lightcurve
is 0.7--2.0 keV. The sinusoid overplotted is the best fit
to any broad modulation, excluding the eclipse itself.}
\end{figure*}  

The spectra of these {\em ASCA} data are also adequately fit by simple 
thermal bremsstrahlung models,
with $\chi^2_r = 0.70$ and 0.78 respectively.
The 2--10~keV fluxes are consistent across the two observations at
$1.79\times10^{-11}$~erg~cm$^{2}$~s$^{-1}$ and 
$1.84\times10^{-11}$~erg~cm$^{2}$~s$^{-1}$, with likely statistical 
uncertainties of order $\sim 1\%$. The fitted column densities too
are consistent with each other with values of 
$N_{\rm H} = 4.1^{+0.3}_{-0.2} \times 10^{22}$~cm$^{-2}$ and  
$N_{\rm H} = 4.4 \pm 0.1 \times 10^{22}$~cm$^{-2}$ respectively. 

\subsection{Archival {\em XMM-Newton} observations}

There are also two archival {\em XMM-Newton} observations of XY Ari, 
neither of which have previously been published either. They comprise 
an observation from 2000 August 26 with exposures of around 16~ksec (MOS) and 
18~ksec (PN) and another from 2001 February 5 with exposures of around 
29~ksec (MOS) and 27~ksec (PN). 
We extracted lightcurves in each case from both the MOS and PN detectors,
in three energy bands. The low energy band corresponds to $\sim 1 - 2$~keV,
the medium energy band is $\sim 2 - 4$~keV, whilst the high band is
$\sim 4 - 10$~keV. In the observation from 2001 February, XY Ari fell
on a chip gap on the MOS1 detector, so the lightcurves are extracted from the
PN and MOS2 only.

Background subtracted lightcurves combined from the MOS and PN 
instruments and folded with an arbitrary phase zero at the white dwarf 
spin period, are shown in Figure 5. To allow comparison between the two
epochs, these lightcurves are each scaled to the count rate from the PN 
detector only. As for the {\em RXTE} data, we show the combined 2--10~keV
data to facilitate comparison with the other observations presented here.
In these pulse profiles the expected double-peaked shapes are 
seen above 2~keV in each observation, but the low energy pulse profile is 
apparently single-peaked at each epoch. As in Figure 1, the best-fitting 
sinusoid with a period equal to half the white dwarf spin period is 
over-plotted on the higher energy lightcurves. The modulation depths, 
listed in Table 1, are consistent between the two observations and between 
the two higher energy bands. 

Figure 6 shows the same lightcurves, folded at the binary orbital
period. In the case of the 2000 observation, there is very little orbital 
phase overlap between the MOS and PN lightcurves, so we show these as 
separate points. In the 2001 observation there is good orbital phase 
overlap between the MOS and PN lightcurves, so in Figure 6 these are 
averaged together. As with Figure 5, these lightcurves are each scaled 
to the count rate of the PN detector only, to allow comparison between
the two epochs, and the two higher energy bands extracted are combined into
a single 2--10~keV lightcurve. As with Figures 2 and 4, we show
the best fit sinusoid to these folded lightcurves (excluding the eclipse) 
overplotted. Here the depths of the modulation, listed in Table 2, are 
significant at the $\sim 4 \sigma$ level or greater in each energy band
during the 2000 observation, but during the 2001 observation the modulation
is not so clearly detected. These {\em XMM-Newton} folded lightcurves show 
much more variation at all phases than the {\em RXTE} data, with hints of a 
structured modulation as a function of orbital phase.

\begin{figure*}[ht]
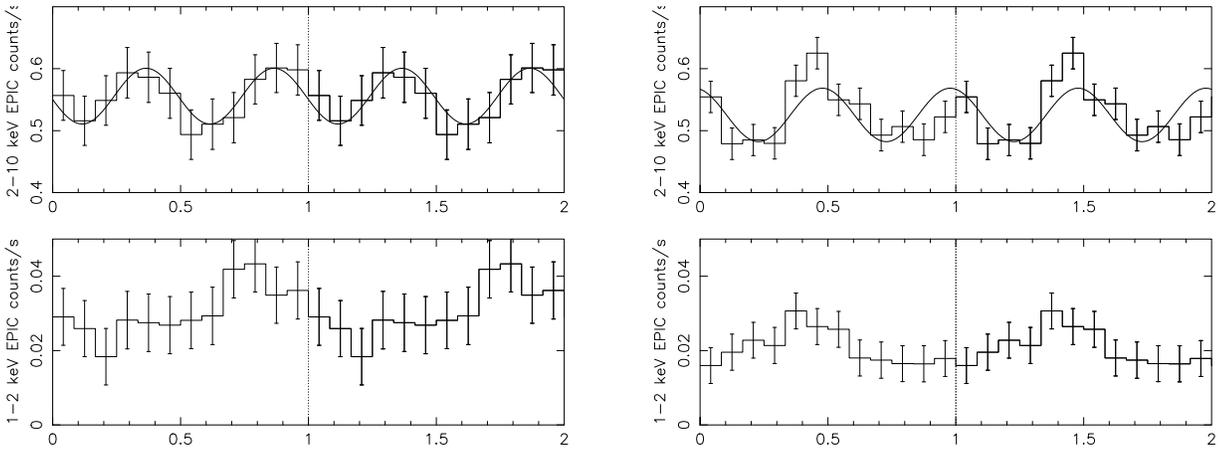

\epsfig{file=7761fig5a.eps, angle=0, width=8.5cm}
\epsfig{file=7761fig5b.eps, angle=0, width=8.5cm}
\caption{The {\em XMM-Newton} combined PN and MOS X-ray lightcurves of XY Ari 
from August 2000 (left) and February 2001 (right) folded at the 206~s white 
dwarf spin period. Count rates are scaled to that of the PN detector 
only in each case. The energy range of the upper lightcurve is 2--10 keV, 
whilst that of the lower lightcurve is 1--2 keV. The sinusoids overplotted 
are the best fit to any double peaked pulse profile.}
\end{figure*}

\begin{figure*}[ht]
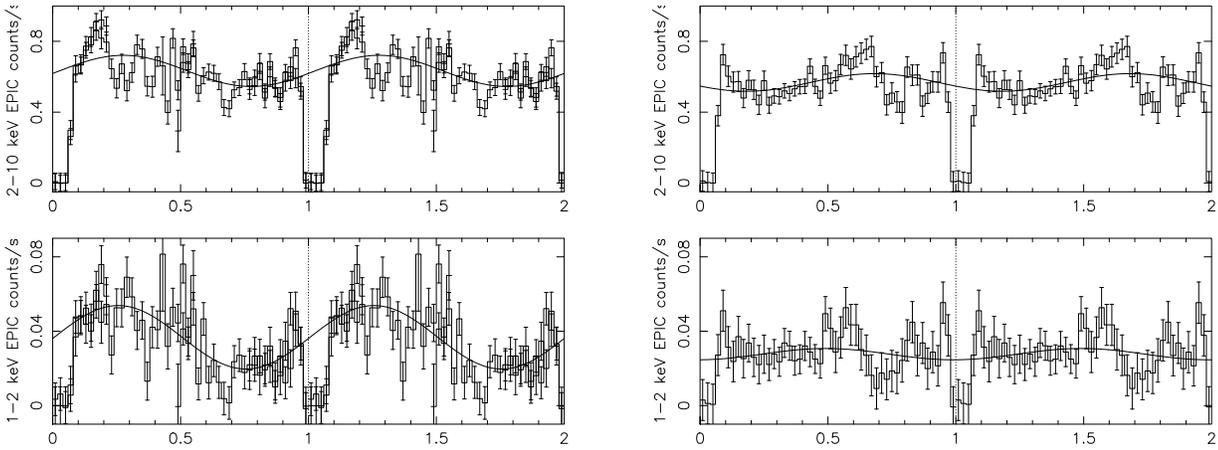

\epsfig{file=7761fig6a.eps, angle=0, width=8.5cm}
\epsfig{file=7761fig6b.eps, angle=0, width=8.5cm}
\caption{The {\em XMM-Newton} combined PN and MOS X-ray lightcurves of XY Ari 
from August 2000 (left) and February 2001 (right)
folded at the 6.06~hr orbital period of the binary. The energy range
of the upper lightcurve is 2--10 keV, whilst that of the lower lightcurve
is 1--2 keV. The sinusoid overplotted is the best fit
to any broad modulation, excluding the eclipse itself. In the 2000 data,
there is little phase overlap between the MOS and PN data, so these 
are plotted separately in this case. The 2001 data are scaled to the 
count rate of the PN only to facilitate comparison.}
\end{figure*}

As with the other observations presented here, the spectra of these 
{\em XMM} data are adequately fit by simple thermal bremsstrahlung models,
with $\chi^2_r = 1.07$ and 1.11 respectively.
The 2--10~keV fluxes are similar for the two observations at
$1.13\times10^{-11}$~erg~cm$^{2}$~s$^{-1}$ and 
$1.20\times10^{-11}$~erg~cm$^{2}$~s$^{-1}$, with likely statistical
uncertainties of order $\sim 1\%$. We note, however, that there is incomplete
orbital phase coverage during the observation in 2000 August, so the flux is
likely to be an under-estimate of the true mean value, as the phase range
sampled includes the eclipse, but not all of the out-of-eclipse region. The 
fitted column densities too are similar between the two observations with 
values of $N_{\rm H} = 4.4 \pm 0.2 \times 10^{22}$~cm$^{-2}$ and  
$N_{\rm H} = 4.8 \pm 0.2 \times 10^{22}$~cm$^{-2}$ respectively. 

\section{Discussion}

In the {\em RXTE} observations of XY Ari presented here, we see two 
unprecedented changes in its behaviour: the previously seen broad orbital
modulation has essentially disappeared, and its double-peaked spin pulse
profile is no longer dominated by phase varying photoelectric absorption.
The archival observations from {\em ASCA} and {\em XMM-Newton} allow 
us to investigate these changes in the context of its long term behaviour.
They show that the extent and phasing of the broad orbital modulation is 
variable on timescales of $\sim$months to years and also that the pulse 
profile varies between double and single peaked, even when the system is not
undergoing an outburst.
The alteration to the orbital modulation implies that changes have occurred
in the outer regions of the accretion disc, whilst the alteration to the
spin pulse profile implies that changes have occurred to the accretion
curtains close to the white dwarf surface. We now consider possible reasons
for each change in behaviour and seek to find a common cause for them.

\subsection{A continued decline from outburst?}

In addition to the changing orbital modulation and spin pulse profile that
we have reported here, another feature that sets XY Ari apart from many other
IPs is the occurence of at least one X-ray outburst, observed 
in July 1996. Although the nature of outbursts in magnetic CVs is debated 
(e.g. Hellier et al. 1997; 2000), if they are similar to those of conventional 
dwarf novae (Cannizzo 1993), it is likely that their recurrence timescale are 
lengthened as a result of the inner disc being truncated by the white dwarf's 
magnetic field (e.g. Matthews et al 2007). The $\sim 30$~yr recurrence period 
of the outbursts in WZ Sge is explained satisfactorily by a magnetic accretor 
model (Matthews et al. 2007), so a $>10$~yr recurrence period in XY Ari 
is not unreasonable.

In order to assess the possibility of a link between a varying mass accretion 
rate and the modulation changes we see, we consider how the observed 
flux and measured column density to XY Ari have changed over the course of 
its various observations. These are summarised in Table 3, where it is clear
that there is little evidence for a significant change in flux following the 
1996 outburst. The average flux before the outburst is $\sim 1.9 \times
10^{-11}$~erg~cm$^{-2}$~s$^{-1}$, and the average flux after the outburst
is  $\sim 1.4 \times 10^{-11}$~erg~cm$^{-2}$~s$^{-1}$, which might reflect a 
$\sim 30\%$ decrease in the mass accretion rate. However,
as all these flux measurements are dependent on the spectral model used
to fit the data in each case, and different instruments were used for
each measurement, the variations are not necessarily significant.
Similarly, the column density, apart from during the outburst itself, has 
remained fairly constant throughout these observations at about 
$4 - 5 \times 10^{22}$~cm$^{-2}$.

Smith et al. (2005) show that the hydrogen column density due to the MBM12
molecular cloud varies across the cloud between 
$2-15 \times 10^{21}$~cm$^{-2}$.
This suggests that a column density of at least $3 \times 10^{22}$~cm$^{-2}$ 
is intrinsic to XY Ari. However, the value we get from a simple bremsstrahlung 
fit to the X-ray spectrum is not necessarily the correct one. 
It is likely that XY Ari, as in most IPs, has a spectrum 
which may be described by two complex, partial covering, absorbers: one for 
the accretion curtain and one for the dip-causing material at the outer edge
of the accretion disc.  Lower accretion rates will reduce both
complex abosorbers, in terms of their column density and/or their covering 
fraction.  Because the molecular cloud reduces our effective bandpass, it is 
not possible to do a complex absorber fit to these faint spectra, so we use a
simple absorbed bremsstrahlung instead.  Such fits can tell us in qualitative
terms when the amount of absorber was higher or lower, but it is not possible
to interpret the column density values obtained in a simple way.

If the change in flux is real then, given that the magnetospheric 
radius is a function of the accretion rate ($R_{\rm mag} \propto 
\dot{M}^{-2/7}$) this might imply an increase in the magnetospheric radius
by about $10\%$. That in turn might allow the lower magnetic pole of the
white dwarf to be more visible, as well as reducing the contribution from the 
impact region at the outer edge of the disc. These two changes together
could then conceivably explain some of the changes in the broad orbital 
modulation and some of the changes in the spin pulse profile that are seen.
However, those modulations do not simply change systematically from before 
to after the outburst. We therefore conclude that
there is no compelling evidence for a significantly altered
X-ray flux or column density between any of the various X-ray observations 
of XY Ari in quiescence. So we must look for other causes of the changes 
in orbital modulation and spin pulse profile.

\begin{table*}
\caption{The X-ray flux and fitted column density 
in various X-ray observations of XY Ari}
\begin{tabular}{llccr} \hline
Mission & Date & 2--10 keV Flux                     & $N_{\rm H}$     & Ref \\
        &      & $/10^{-11}$~erg~cm$^{-2}$~s$^{-1}$ & $/10^{22}$~cm$^{-2}$ & \\ \hline
{\it Ginga}   & 1989 Aug & 2.5   & $3 - 16$ & Kamata \& Koyama 1993 \\
{\it ASCA}    & 1995 Aug & 1.79  & $3.9 - 4.4$ & this work \\
{\it ASCA}    & 1996 Jan & 1.84  & $4.3 - 4.5$ & this work \\
{\it RXTE}    & 1996 Jul & 1.5   & $\sim 6 - 12$ & Hellier et al 1997\\
{\it RXTE}    & 1996 Jul (o/b) &  5.5 & $\sim 40$ & Hellier et al 1997\\
{\it Chandra} & 2000 Jul & 1.8   & $4.6 - 5.6$ & Salinas \& Schlegel 2004 \\
{\it XMM}     & 2000 Aug & 1.13  & $4.2 - 4.6$ & Nb. incomplete orbital coverage\\
{\it XMM}     & 2001 Feb & 1.20  & $4.6 - 5.0$ & this work \\
{\it RXTE}    & 2005 Nov & 1.45  & $4.1 - 6.4$ & this work \\
{\it RXTE}    & 2006 Jan & 1.32  & $5.4 - 8.1$ & this work \\ \hline
\end{tabular}
\end{table*}

\subsection{A changing orbital modulation}

The broad orbital modulation seen from XY Ari was prominent when the object 
was observed by {\em Ginga}, {\em ASCA}, {\em Chandra} and {\em XMM-Newton}
between 1989 and 2000. By the time of the second {\em XMM-Newton} observation
in 2001, the broad modulation was less apparent with a significant
decrease in the low energy  modulation depth.
In the most recent observations, with {\em RXTE}, the broad modulation
is no longer detected with any significance. We note also that the phase of
minimum of the broad modulation appears to drift significantly from one
observation to another when the modulation itself is prominent. 

Since the broad modulation is believed to be due to absorption by an
azimuthally localised structure near the edge of an accretion disc, its 
disappearance suggests one of three possibilities: the absorbing structure 
may have diminished, extended to all azimuths, or otherwise 
moved out of our line of sight, at the time of these observations. As noted 
earlier, a similar effect has previously been seen in AO Psc, 
where an orbital modulation was seen by {\em EXOSAT} (1983 and 1985), 
{\em Ginga} (1990), {\em ROSAT} (1994) and {\em ASCA} (1994) but not by 
{\em RXTE} (1997); whereas  for V1223 Sgr an orbital modulation was seen by 
{\em EXOSAT} (1983 and 1984), not seen by {\em Ginga} (1991) or {\em ROSAT}
(1994), then seen again by {\em ASCA} (1994) and {\em RXTE} (1998) (Parker, 
Norton \& Mukai 2005). In that earlier paper, we preferred the third of the 
possibilities above as the likely cause and suggested that the varying
appearence of such orbital modulations may be due to the precession of a
tilted accretion disc. This might explain both the drift of the phase of 
minimum and the ultimate disappearence of the modulation on long timescales
as the extended bulge at the outer edge of the accretion disc moves out
of our line of sight to the X-ray source.

One IP, TV Col, is known to have a 4~d photometric variation which is 
assumed to represent a disc precession period (Barrett, O'Donoghue \& 
Warner 1988), and similar effects are seen in several X-ray binaries. The
Low Mass X-ray Binary system  4U1916--053 displays X-ray `dips' as its 
tilted disc periodically obscures the central X-ray source (Homer et al 2001) 
and `superorbital' X-ray periods have been noted in systems such as Her X-1, 
SS433 and LMC X-4, which are also explained by precessing tilted or warped 
accretion discs (e.g. Clarkson et al 2003). 

Smoothed Particle Hydrodynamic simulations of accretion discs irradiated by 
a central X-ray source have been shown to induce twists or warps into the 
discs, which tilt and then precess with periods of a few days (Foulkes, 
Haswell \& Murray 2006). These authors ran simulations for systems with a 
range of mass ratios and X-ray luminosities and showed that radiation-driven
warping occurred in all cases. For systems with a relatively high mass ratio,
such as XY Ari with $0.43 < q < 0.71$, the entire disc tilts out of the 
orbital plane, and then precesses. Foulkes et al (2006) showed that the rate 
of precession is faster, and the tilt is larger, for higher X-ray 
luminosities. Whilst the mass ratio of XY Ari fulfils the necessary criterion, 
its X-ray luminosity ($L_{\rm X} / L_{\rm Edd} \sim 5 \times 10^{-6}$) is 
significantly lower than any system they modelled and is unlikely to be 
strong enough to drive the warp (Foulkes, private communication).

However, Foulkes et al. did not take into account any interaction between
the compact object's magnetic field and the accretion disc. It is conceivable
that, in a system such as XY Ari where the white dwarf's magnetic field is
inclined with respect to the disc plane, the field lines attaching to the 
inner (truncated) disc edge may themselves serve to induce a warp in the
disc as they rotate with respect to the disc material. In this case, such 
warps may be common amongst IPs and we encourage the 
construction of simulations to investigate this possibility.

The fact that we see no evidence for changes in the X-ray orbital 
modulation of XY Ari on the $\sim 1$ month timescale which corresponds to the 
gap between the two halves of the {\em RXTE} observation suggests that the 
disc precession period in XY Ari may be rather long -- perhaps of order months.
Although the presence of the broad modulation would be expected to come and 
go on the timescale of the precession period, we note that the broad 
modulation itself would actually occur on the beat period
between the precession and orbital periods. Such a modulation is indeed seen
in TV Col at a period of 5.2~h, which is the beat between the 5.5~h orbital
period and the 4~d retrograde disc precession period (Hellier 1993). However, 
with a precession period of (say) $>100$~d, the beat period in XY Ari would 
only differ from the orbital period by $<0.25\%$ and so would be essentially 
indistinguishable from it in this case. Hence we cannot use the broad 
modulation (when present) to measure the precession period.

\subsection{A changing spin pulse profile}

If the changing orbital modulation in XY Ari really is due to a precessing
disc, the next question to address is whether this can 
explain the observed variations in the spin pulse profile too.

On its discovery by {\em Ginga} in 1989, XY Ari exhibited a double-peaked 
pulse profile which showed a decreasing modulation depth with increasing 
energy, up to 10 keV (Kamata \& Koyama 1993). Such a pulse profile, which 
is seen in several IPs, has been 
interpreted as indicating emission from two accreting poles, both of 
which are visible (Norton et al 1999; Hellier, Mukai \& Beardmore 1997).
The observed variation in modulation depth with energy is as expected if 
the modulation is produced by phase-varying photoelectric absorption close to 
the surface of the white dwarf in the accretion curtains. However, we note 
that the lowest energy pulse profile presented by Kamata \& Koyama (1993) 
(i.e. below $\sim 5$~keV) shows a structure in which the `second' peak is 
significantly dimished, indicating that the absorbing structure is different 
for the two accretion poles in this case. When a similar double-peaked
pulse profile was seen by {\em RXTE} observations of XY Ari both before and
after its 1996 outburst, Hellier et al (1997) explained the low amplitude of
the pulse by attributing it to an asymmetry between the emission at the
two poles, without which cancellation would occur and a constant flux would 
result.  The only other previously published observations of XY Ari in 
quiescence are those obtained by {\em Chandra} in 2000. Salinas \& Schlegel 
(2004) show that the pulse profile in these observations is rather noisy, 
and although there is evidence for a varying hardness ratio with phase, the 
profile shape is essentially single-peaked.

During the 1996 outburst observed by {\em RXTE}, the modulation depth of 
XY Ari's spin pulse was greatly enhanced 
and the profile changed from its former double-peaked structure to a smoothly 
sinusoidal profile (Hellier et al. 1997). The interpretation placed on this 
was that the inner disc extended closer to the white dwarf during outburst, 
so hiding the lower magnetic pole. After the outburst, the truncation radius 
of the disc increased once more, revealing the lower pole and so allowing the 
`usual' double-peaked pulse profile, dominated by absorption effects at both 
poles, to return.

In contrast to these previous observations, the spin pulse profiles seen 
in our {\em RXTE} observations of XY Ari from 2005/2006 are clearly 
double-peaked, but show no evidence for energy dependence, and therefore 
indicate that the dominant cause of the spin modulation at this time is not 
photoelectric absorption, but must be essentially geometric.

We further note the different pulse profiles seen during the {\em ASCA} and
{\em XMM-Newton} observations which we report here for the first time. 
Although the {\em ASCA} observations preceded Hellier et al's 1996
{\em RXTE} observations of XY Ari in quiescence by only a few months, the 
pulse profile was either single-peaked or absent in those observations. 
Furthermore, although the {\em XMM-Newton} observations followed that
with {\em Chandra} by only a few months, the pulse profile observed there is 
clearly double-peaked above 2~keV. These changes are summarised in Table 1.

It is apparent from the above discussion that the visibility of the two
accretion poles in XY Ari can change on a timescale of months or less. 
One way of accomplishing this may be for the white dwarf itself to precess,
such that the magnetic field axis presents a varying inclination angle to the
orbital plane. Tovmassian, Zharikov \& Neustroev (2007) have recently 
presented evidence for such behaviour in the probable IPs FS Aur and 
HS~2331+3905. The rapidly spinning white dwarfs in these systems (with
spin periods of $\sim 100$~s and 67.2~s respectively) have precession periods 
of just a few hours. Schwarzenberg-Czerny (1992) estimated that the 
white dwarf precession period would be $\sim 10^4 - 10^5 \times$ the white
dwarf spin period in typical IPs, implying a precession period of $\sim 24 - 
240$~d in XY Ari, although the numerical calculations of Leins et al (1992)
presented graphically by Tovmassian et al (2007) suggest a shorter precession
period of just a few days in this case. However, we note that we see no change
in the X-ray lightcurves between the two {\em RXTE} visits reported here 
(separated by $\sim 50$~d) nor do we see any change on a timescale of 
$\sim 4$~d by subdividing the visits (although the signal-to-noise ratio is
then greatly reduced). 

However, rather than postulate a precessing white dwarf {\em and} a precessing
accretion disc, it is possible that the disc itself could hide the lower 
accreting pole to varying degrees at different times, if it is indeed tilted.
Occasional hiding of the lower pole by the inner edge of the disc could
certainly explain the appearance of a single peaked pulse profile (as seen 
in the {\em ASCA} and {\em Chandra} observations), using the same model as 
suggested by Hellier et al (1997) to explain the single-peaked pulse
profile seen in outburst. As the top of the accretion curtains will be 
anchored at the inner edge of the disc, we suggest that a precessing disc 
could also cause the absorbing material in the curtains to be largely removed
from our line of sight to the emission regions, so giving rise to the 
energy-independent pulse profiles we see in the latest {\em RXTE} observations.

\section{Conclusion}

We have demonstrated that the {\em RXTE} observations of XY Ari from November
2005 and January 2006 reveal its broad X-ray orbital modulation to have 
disappeared. At the same time, its spin pulse profile displays no evidence
of phase-varying photoelectric absorption. Both of these features are in 
contrast to earlier observations of the source dating back to fifteen years
earlier. However, observations with other X-ray satellites over the 
intervening years have revealed that the broad orbital modulation drifts 
in both phase and modulation depth, and that the spin pulse profile varies 
too in both shape and significance. However, there is no evidence for a 
systematic change in X-ray flux or column density of XY Ari (as the disc 
presumably settles into quiescence following its 1996 outburst) to explain 
the observed changes in behaviour. Instead we suggest that a precessing, 
tilted accretion disc may cause both the changes in the broad orbital 
modulation and the changes in the pulse profile. At its outer edge, the 
precessing disc will move the absorbing bulge in orbital phase and may remove 
it from our line of sight entirely. At its inner edge, the precessing disc 
will alter our view through the accretion curtains that are anchored there, 
and may also hide our view of the lower pole. The tilt and precession of the
disc may be induced by the action of the inclined, rotating magnetic field
lines at the disc's inner edge; it is unlikely to be driven by radiation
from the central source.

We conclude by noting that since XY Ari is the only deeply eclipsing 
IP, in this source we are probing matter which is less 
than $6^{\circ}$ above the orbital plane. Other IPs may also 
possess precessing, tilted accretion discs. However, the influence of such a 
structure is more apparent in XY Ari than in other systems where we see an 
X-ray orbital modulation, because in those cases our line of sight is up to 
$30^{\circ}$ above the orbital plane. With the additional characteristic of 
showing outbursts on a timescale which may be $\sim 10$~yr, XY Ari therefore 
allows us a unique view of the accretion disc structure amongst magnetic 
cataclysmic variables.

\begin{acknowledgements}

The research described here has made use of data from NASA's 
High Energy Astrophysics Science Archive Research Center. Astrophysics
Research at the Open University is supported by PPARC Rolling Grant
PP/D000963/1. We thank Dr. S. Foulkes for useful discussions and the
anonymous referee for several interesting suggestions.

\end{acknowledgements}


\begin{thebibliography}{}

 \bibitem[\protect\citename{}]{} Allan, A., Hellier, C., Ramseyer, T.F. 1996,
MNRAS, 282, 699

 \bibitem[\protect\citename{}]{} Barrett, P., O'Donoghue, D., Warner, B. 1988,
MNRAS, 233, 759

 \bibitem[\protect\citename{}]{} Beardmore, A.P., Mukai, K., Norton, A.J.,
Osborne, J.P., Hellier, C. 1998, MNRAS, 297, 337

 \bibitem[\protect\citename{}]{} Cannizzo, J. 1993, ApJ, 419, 318

 \bibitem[\protect\citename{}]{} Clarkson, W.I., Charles, P.A., Coe, M.J.,
Laycock, S. 2003, MNRAS, 343, 1213

 \bibitem[\protect\citename{}]{} Ezuka, H., Ishida, M. 1999, ApJS, 120, 277

 \bibitem[\protect\citename{}]{} Foulkes, S.B., Haswell, C.A., Murray, J.R.
2006, MNRAS, 366, 1399

 \bibitem[\protect\citename{}]{} Halpern, J.P., Patterson, J. 1987, ApJL,
312, L31

 \bibitem[\protect\citename{}]{} Hellier, C. 1993, MNRAS, 264, 132

 \bibitem[\protect\citename{}]{} Hellier, C. 1997, MNRAS, 291, 71

 \bibitem[\protect\citename{}]{} Hellier, C., Mukai, K., Beardmore, A.P.
1997, MNRAS, 292, 397

 \bibitem[\protect\citename{}]{} Hellier, C., Kemp, J., Naylor, T., 
Bateson, F.M., Jones, A., Overbeek, D., Stubbings, R., Mukai, K. 2000,
MNRAS, 313, 703 

 \bibitem[\protect\citename{}]{} Hellier, C., Harmer, S., Beardmore 2004, 
MNRAS, 349, 710

\bibitem[\protect\citename{}]{} Homer, L., Charles, P.A., Hakala, P., 
Muhli, P., Shih, I.-C., Smale, A.P., Ramsay, G. 2001, MNRAS, 322, 827

 \bibitem[\protect\citename{}]{} Kamata, Y., Tawara, Y., Koyama, K. 1991,
ApJL, 379, L65

 \bibitem[\protect\citename{}]{} Kamata, Y., Koyama, K. 1993, ApJ, 405, 307

 \bibitem[\protect\citename{}]{} Koyama, K., et al. 1991, ApJ, 377, 240

 \bibitem[\protect\citename{}]{} Leins, M., Soffel, M.H., Lay, W., Ruder, H. 
1992, A\&A, 261, 658

 \bibitem[\protect\citename{}]{} Littlefair, S., Dhillon, V., Marsh, T.R.
2001, MNRAS, 327, 669

 \bibitem[\protect\citename{}]{} Matthews, O.M., Speith, R., Wynn, G.A.,
West, R.G. 2007, MNRAS, 375, 105

 \bibitem[\protect\citename{}]{} Norton, A.J., Beardmore, A.P., Allan, A., 
Hellier, C. 1999, A\&A, 347, 203

 \bibitem[\protect\citename{}]{} Norton, A.J., Wynn, G.A., Somerscales, R.V.
2004, ApJ, 614, 349

 \bibitem[\protect\citename{}]{} Parker, T.L., Norton, A.J., Mukai, K. 2005,
A\&A, 439, 213

 \bibitem[\protect\citename{}]{} Patterson, J. 1994, PASP, 106, 209

 \bibitem[\protect\citename{}]{} Patterson, J., Halpern, J.P. 1990, ApJ, 
361, 173

 \bibitem[\protect\citename{}]{} Salinas, A., Schlegel, E.M. 2004, AJ, 
128, 1331

 \bibitem[\protect\citename{}]{} Schwarzenberg-Czerny, A. 1992, A\&A, 260, 268


 \bibitem[\protect\citename{}]{}Smith, R.K., Edgar, R.J., Plucinsky, P.P., 
Wargelin, B.J., Freeman, P.E., Biller, B.A. 2005, ApJ, 623, 225

 \bibitem[\protect\citename{}]{} Takano, S. et al 1989, IAUC 4745

 \bibitem[\protect\citename{}]{} Terada, Y., Ishida, M., Makishima, K. 2004,
PASJ, 56, 533

 \bibitem[\protect\citename{}]{} Tovmassian, G.H., Zharikov, S.V., Neustroev, 
V.V. 2007, ApJ, 655, 466

 \bibitem[\protect\citename{}]{} Warner, B. 1995, Cataclysmic Variable Stars,
Cambridge Astrophysics Series (Cambridge, New York: Cambridge University
Press)

 \bibitem[\protect\citename{}]{} Watson, M.G., King, A.R., Osborne, J.P.
1985, 212, 917

\end{thebibliography}
\end{document}